\documentclass[sigconf,nonacm,balance=false]{acmart}
\usepackage{amsmath,booktabs,graphicx}
\setcopyright{none}
\acmDOI{}
\acmISBN{}
\acmConference[]{}{}{}
\acmYear{}
\copyrightyear{}
\graphicspath{{./}{figures/}}
\newcommand{\bp}{\,bp}

\title[Herding or Racing?]{Herding or Racing? Decomposing Cross-Account Order-Flow Persistence in On-Chain Markets}
\titlenote{Accepted at the workshop ``AI-Driven Market Microstructure in Centralized and Decentralized Finance'' at the 7th ACM International Conference on AI in Finance (ICAIF~'26), Milan, Italy, November 2026.}
\author{Yichong Bai}
\email{byc@pku.org.cn}
\affiliation{\institution{The Hong Kong Polytechnic University}\city{Hong Kong SAR}\country{China}}

\begin{document}
\begin{abstract}
Most of the autocorrelation in trade signs on decentralized exchanges comes from pairs of trades sent by different wallets. For Uniswap~v3, this has been read as herding dominating order splitting---the reverse of what equity markets show. We show that this herding share mixes two quantities: the probability $p$ that consecutive orders come from the same account, which is small when thousands of wallets are trading, and the ratio $\kappa$ of cross-account to same-account sign correlation. For any $p$, $\kappa$ equals one under the public-signal benchmark of T\'oth et al.\ and zero under independent order splitting. On Uniswap~v3 USDC--WETH (one week each in 2023 and 2024), the herding share is 0.65--0.86, but $\kappa$ is only 0.05--0.10. On a fully on-chain EVM limit order book, the herding share is similar (0.61--0.87), but $\kappa$ is 0.72--0.76; among Uniswap swaps above \$100k, it is 0.34--0.52. Where $\kappa$ is high, cross-account co-movement is mostly a race for the price on a centralized exchange. It is concentrated in orders a few blocks apart, and a cross-fitted gradient-boosted filter that sees only Binance data removes about three quarters or more of it. This race flow accounts for most of the losses of liquidity providers. We recommend reporting $\kappa$ alongside the herding share and modeling on-chain order flow as splitting traders plus arbitrageurs who respond to the centralized-exchange price.
\end{abstract}
\maketitle

\section{Introduction}
Market-order signs are persistent. In equity markets, this persistence is mostly attributed to order splitting, as investors slice large metaorders into many small trades~\cite{lillo2004longmemory,lillo2005theory}. The decisive evidence comes from splitting the sign autocorrelation into pairs of orders placed by the same agent and pairs placed by different agents~\cite{toth2015persistent}. On the London Stock Exchange, about three quarters of the lag-one correlation comes from the same member firm, and account-level data from Tokyo confirm the splitting mechanism quantitatively~\cite{sato2023inferring}; see~\cite{lillo2023orderflow} for a review. T\'oth et al.\ also caution that a large cross-agent term is necessary but not sufficient for herding.

Decentralized exchanges (DEXs) allow the same decomposition at scale, since every on-chain order carries the address that sent it. Di Nosse, Gatta, Lillo and Jaimungal~\cite{dinosse2026deviations} study the 24 most active Uniswap~v3 pools in 2023--2024 and find the opposite of the equity result: the cross-wallet (``herding'') component dominates the autocorrelation of swap signs. The result holds up when sandwich attacks are removed and routed swaps are aggregated. They connect it to the herding effect, ``the tendency of investors to replicate the trades of others'', and note that cryptocurrency prices show ``many strong price trends that can be attributed to herding effects''. They list it among the main ways decentralized finance (DeFi) departs from the stylized facts of traditional markets.

We ask what this herding share actually measures, and make three contributions.

\emph{Measurement.} The herding share $h$ is an exact function of two quantities: the probability $p$ that two consecutive orders come from the same account, and the ratio $\kappa$ of the sign correlation between orders of different accounts to that between orders of the same account. Under either standard benchmark, $\kappa$ does not depend on $p$: it is one when traders respond to a common public signal and zero when accounts split orders independently. When thousands of wallets trade, $p$ is only a few per cent, so $h$ can be large even if $\kappa$ is close to zero. On the pool that Di Nosse et al.\ use as their main example, $h$ is 0.65--0.86 while $\kappa$ is 0.05--0.10: pair by pair, small swaps behave like equity order flow. On a fully on-chain limit order book built on the Ethereum Virtual Machine (EVM), $h$ is similar, but $\kappa\approx0.75$.

\emph{Mechanism.} Where $\kappa$ is high---on the order book and among Uniswap swaps above \$100k---cross-account co-movement is mostly a race for the price on a centralized exchange (CEX). A cross-fitted gradient-boosted classifier that sees only Binance data removes about three quarters or more of it, and on the order book delaying its inputs by half a second makes little difference. The co-movement is concentrated in orders a few blocks apart. On the order book it is strong within a block and lasts for about ten blocks; among large Uniswap swaps it is weaker or negative within a block and concentrated in the next few blocks.

\emph{Economics.} Liquidity providers pay mostly for the race. On the order book, orders in a cross-account race make up 54--74\% of volume and carry 70--91\% of makers' losses. On Uniswap, after the pool fee, liquidity providers (LPs) lose 4--5\bp{} on the third of volume most aligned with Binance and earn about 5\bp{} on the least aligned third.

\paragraph{Related work.} Beyond the order-flow literature above, our work connects to studies of maximal extractable value (MEV) and transaction ordering~\cite{daian2020flash}, CEX--DEX arbitrage and the losses it imposes on liquidity providers~\cite{milionis2022automated,milionis2025automated,heimbach2024sp,wu2025measuring}, informed traders who pay for priority on DEXs~\cite{capponi2026price}, and latency arbitrage on stale quotes in traditional markets~\cite{budish2015high,foucault2017toxic,aquilina2022quantifying}. Recent identity-level studies of on-chain order books show that taker toxicity is a persistent wallet attribute~\cite{zhai2026identity}, reconstruct address-level metaorders~\cite{barone2026sunshine}, and release full message data~\cite{albers2026openbook}. Client identity also predicts toxic flow in foreign exchange~\cite{cartea2026toxic}, and Binance and a few other venues lead price formation across bitcoin derivatives markets~\cite{albers2021fragmentation}. None of these decomposes sign persistence against a CEX signal. Methodologically, we borrow cross-fitting from double machine learning~\cite{chernozhukov2018double} to use a flexible learner~\cite{friedman2001greedy,sklearn2011} as a measurement instrument.

\section{Method}
\subsection{The herding ratio \texorpdfstring{$\kappa$}{kappa}}
Let $\epsilon_n\in\{-1,+1\}$ be the sign of the $n$-th order ($+1$ buys the base asset), $a_n$ its account, and $x_n=\epsilon_n-\bar\epsilon$. Following~\cite{toth2015persistent,dinosse2026deviations}, split the lag-$\ell$ autocovariance $C(\ell)$, the mean of $x_nx_{n+\ell}$, into $C_{\mathrm{same}}(\ell)$ and $C_{\mathrm{other}}(\ell)$ according to whether $a_n=a_{n+\ell}$. Let $p(\ell)$ be the share of same-account pairs, and let $\rho_{\mathrm{same}}=C_{\mathrm{same}}/p$ and $\rho_{\mathrm{diff}}=C_{\mathrm{other}}/(1-p)$ be the mean products within each type of pair. The herding share is then
\begin{equation}\label{eq:share}
 h(\ell)=\frac{C_{\mathrm{other}}}{C}=\frac{(1-p)\,\kappa}{p+(1-p)\,\kappa},\qquad \kappa(\ell)=\frac{\rho_{\mathrm{diff}}}{\rho_{\mathrm{same}}}.
\end{equation}
Two benchmarks anchor $\kappa$. In the public-signal model of~\cite{toth2015persistent}, randomly selected traders respond to a common signal, every pair has the same correlation, and $\kappa=1$. With independently drawn account labels, $p=\sum_iP_i^2$, where $P_i$ is account $i$'s share of orders; this recovers their benchmark for the same-account share $S=1-h$. If accounts split orders independently of one another, then $\rho_{\mathrm{diff}}=0$ and $\kappa=0$.

The herding share inherits $p$, which depends on how many accounts trade and how their orders cluster in time. Under both benchmarks, $\kappa$ does not, and unlike the comparison of $S$ with $\sum_iP_i^2$, it does not assume that account labels arrive independently. In data, $\kappa$ can still vary with the mix of accounts and with the size threshold, and it can exceed one. Using account-specific means as in~\cite{dinosse2026deviations} instead of $\bar\epsilon$ changes $\kappa$ by at most 0.04 and $h$ by at most 0.02 in our samples.

\subsection{A learned common-signal filter}
To measure how much of the co-movement a public signal explains, we estimate $\hat s_n=\mathrm{E}[\epsilon_n\mid X_n]$ from features $X_n$ observed before the order, and decompose the residual $u_n=\epsilon_n-\hat s_n$ in the same way. The share of cross-account covariance that the filter removes is $1-\rho^{u}_{\mathrm{diff}}/\rho^{x}_{\mathrm{diff}}$. The learner is histogram-based gradient boosting~\cite{friedman2001greedy,sklearn2011} with 300 iterations, learning rate 0.05, 31 leaves, an $L_2$ penalty of 1, and a minimum leaf size of 200 orders (order book) or 50 swaps (Uniswap). We cross-fit~\cite{chernozhukov2018double}: the sample is split by the parity of the clock hour, and each prediction comes from a model trained on the other half, so the filter cannot absorb in-sample noise into the residual statistics.

$X_n$ contains only Binance data, read from a 100\,ms grid:
\begin{itemize}
\item returns over eight windows covering the 30\,s before the order on the order book, or seven windows covering the 5\,min before the slot on Uniswap;
\item signed trade volume over three windows (order book) or two (Uniswap);
\item realized volatility over 60\,s (order book) or 5\,min (Uniswap).
\end{itemize}
The latest input is the grid cell containing the block (order book) or slot (Uniswap) time, so it can include up to 100\,ms of later Binance data. As a leakage control, we refit the order-book filter with all inputs delayed by 0.5\,s, which leaves only data from at least 0.4\,s before the block. A variant adds the gap between the venue's price and the Binance-implied fair value, which is what arbitrageurs trade on but is not exogenous to the venue; on the order book it uses the gap at four lags and the spread. The fair value is the Binance price times a slowly varying basis: the 300\,s rolling median of the venue-to-Binance midpoint ratio on the order book, and the median pool-to-Binance price ratio over the previous 400 swaps on Uniswap.

\section{Data}
\paragraph{On-chain order book.} The venue is a central limit order book implemented as EVM smart contracts on a chain with sub-second blocks. Every order and fill is emitted as a contract event that carries the account identifier. We study its BTC perpetual contract over two samples: a main sample of 116 hours in late September and early October 2026, and an earlier sample of 72 hours in late August 2026. We use filled taker orders of at least \$10. Smaller orders are 37\% and 30\% of orders but 0.2\% and 0.6\% of volume; including them gives $\kappa=0.77$ and $h=0.76$ in the main sample. The main sample has 94{,}469 orders from 302 accounts and \$49.5M of volume; the earlier sample has 142{,}659 orders from 457 accounts and \$69.3M. Taking is concentrated in the main sample, where two accounts supply 75\% of volume, but not in the earlier one (24\%).

In the main sample, block times come from the first receipt of each block's update on the venue's public book feed, corrected for the median delivery delay; they differ from block times observed through a public node subscription by about 45\,ms in median absolute terms. For the earlier sample, we interpolate the chain's whole-second block timestamps and shift them by 0.65\,s, their median offset from observed block times on days when both are available. The leader is Binance's USD-margined BTCUSDT perpetual: best quotes on a 100\,ms grid in the main sample, and last trade prices from Binance's public archive in the earlier one (on a common day the two agree to 0.02\bp). Makers' markouts are measured at 60\,s, before trading fees and gas, against the venue's midpoint (main sample) or the Binance price times a five-minute rolling median basis estimated from fill prices (earlier sample).

\paragraph{Uniswap v3.} We use the USDC--WETH 0.05\% pool, the main example of~\cite{dinosse2026deviations}. We extract all swap events for 3--9 June 2024 and 5--11 June 2023 from public Ethereum nodes, and take the transaction sender as the wallet, as they do. Swaps are signed $+1$ when they buy WETH, and we keep swaps of at least \$10. That leaves 37{,}336 swaps from 17{,}982 wallets (\$1.12bn) in 2024, and 31{,}843 swaps from 13{,}464 wallets (\$1.93bn) in 2023. Sandwich and echo patterns detected with their algorithm are 0.35--0.62\% of swaps and 1.9--6.5\% of dollars; Table~\ref{tab:decomp} keeps them, and we report their effect below. The leader is Binance's ETHUSDT perpetual, aligned to the slot timestamp. LP markout is the execution price relative to the Binance-implied fair value 60\,s later. The execution price recorded in the swap event includes the 5\bp{} pool fee, so this markout is the LPs' result after the fee. All intervals are 95\% hour-block bootstrap intervals with 2{,}000 resamples.

\section{Results}
\begin{table*}[t]
\caption{Decomposition of the lag-one autocorrelation of order signs. $N$: orders of at least \$10; $p$: share of consecutive pairs from the same account; $\rho_{\mathrm{same}}$, $\rho_{\mathrm{diff}}$: mean products of demeaned signs within each type of pair; $\kappa=\rho_{\mathrm{diff}}/\rho_{\mathrm{same}}$; $h$: herding share of $C(1)$; $S/\sum P_i^2$: same-account share relative to the public-signal benchmark of~\cite{toth2015persistent} (about 12 on the London Stock Exchange). Brackets show 95\% hour-block bootstrap intervals.}
\label{tab:decomp}
\small
\begin{tabular}{@{}lrrrrrccr@{}}
\toprule
Sample & $N$ & Accounts & $p$ & $\rho_{\mathrm{same}}$ & $\rho_{\mathrm{diff}}$ & $\kappa$ & $h$ & $S/\sum P_i^2$\\
\midrule
Order book, main sample & 94{,}469 & 302 & 0.331 & 0.831 & 0.634 & 0.76 [0.73, 0.82] & 0.61 [0.57, 0.64] & 2.2 \\
Order book, earlier sample & 142{,}659 & 457 & 0.098 & 0.953 & 0.689 & 0.72 [0.70, 0.74] & 0.87 [0.86, 0.88] & 4.5 \\
\midrule
Uniswap 2024, all swaps & 37{,}336 & 17{,}982 & 0.025 & 0.810 & 0.037 & 0.05 [0.03, 0.06] & 0.65 [0.55, 0.73] & 199.8 \\
\quad swaps $\geq$ \$10k & 9{,}411 & 2{,}753 & 0.042 & 0.883 & 0.167 & 0.19 [0.15, 0.23] & 0.81 [0.77, 0.84] & 26.8 \\
\quad swaps $\geq$ \$100k & 3{,}816 & 526 & 0.104 & 0.761 & 0.257 & 0.34 [0.25, 0.43] & 0.75 [0.69, 0.79] & 9.6 \\
\midrule
Uniswap 2023, all swaps & 31{,}843 & 13{,}464 & 0.016 & 0.732 & 0.072 & 0.10 [0.08, 0.12] & 0.86 [0.83, 0.89] & 63.3 \\
\quad swaps $\geq$ \$10k & 12{,}499 & 3{,}123 & 0.035 & 0.787 & 0.226 & 0.29 [0.24, 0.34] & 0.89 [0.87, 0.91] & 9.0 \\
\quad swaps $\geq$ \$100k & 5{,}215 & 748 & 0.059 & 0.667 & 0.346 & 0.52 [0.43, 0.62] & 0.89 [0.87, 0.91] & 4.4 \\
\bottomrule
\end{tabular}
\end{table*}

\subsection{Same herding share, different \texorpdfstring{$\kappa$}{kappa}}
Table~\ref{tab:decomp} reports the decomposition. On Uniswap, the herding share is 0.65 in 2024 and 0.86 in 2023, confirming the finding of~\cite{dinosse2026deviations} for their example pool. Yet $\kappa$ is only 0.05 and 0.10. Consecutive swaps from the same wallet agree with a correlation of 0.73--0.81, while consecutive swaps from different wallets agree with 0.04--0.07. The herding share is high because 97--98\% of consecutive swaps come from different wallets. Measured against the public-signal benchmark, the same-wallet share is 63--200 times $\sum_iP_i^2$, against about 12 times for the London Stock Exchange~\cite{toth2015persistent}. Removing sandwich and echo patterns lowers $\kappa$ to 0.04 and 0.07 and raises the same-wallet correlation to 0.90 and 0.96. Pair by pair, small DEX flow is therefore dominated by splitting, as equity order flow is.

On the order book, the herding share is similar (0.61 and 0.87), but $\kappa$ is 0.76 and 0.72. Orders from different accounts are nearly as correlated as orders from the same account, close to the public-signal benchmark. Uniswap swaps of \$100k or more, which carry 76--82\% of the pool's volume, lie in between ($\kappa=0.34$--$0.52$); sandwich and echo patterns change these values by at most 0.02. The size threshold matters in opposite directions. On Uniswap, $\kappa$ rises with it (Table~\ref{tab:decomp}). On the order book it falls, to 0.73 for orders of at least \$100 and 0.51 for the 22\% of orders above \$500.

\begin{figure}[t]
\centering
\includegraphics[width=\columnwidth]{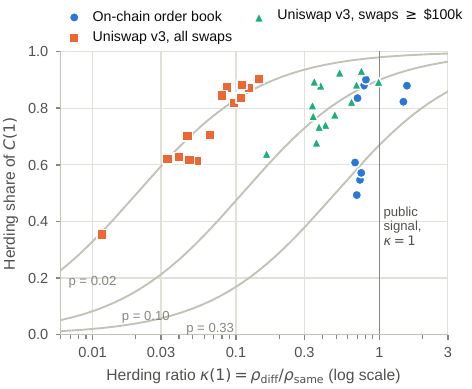}
\Description{Scatter plot of the daily herding share against the herding ratio kappa on a log scale. Order-book days lie between kappa 0.7 and 1.6, Uniswap days for all swaps between 0.01 and 0.15, and days for large Uniswap swaps in between. Grey curves show the identity for p equal to 0.02, 0.10 and 0.33.}
\caption{Daily herding share against the herding ratio $\kappa$ (lag one). Grey curves show Eq.~\eqref{eq:share} for three values of $p$. The herding share confounds $\kappa$ with $p$. $\kappa$ separates all Uniswap swaps from the order book; large Uniswap swaps lie in between.}
\label{fig:kappa}
\end{figure}

Figure~\ref{fig:kappa} shows the daily values. Each lies on the curve of Eq.~\eqref{eq:share} for its own $p$, and the same herding share arises with values of $\kappa$ that differ by more than an order of magnitude. On the last two days of the main sample, $\kappa$ exceeds one (1.47 and 1.56), because the same-account correlation falls to 0.41 while the cross-account correlation stays at 0.61--0.65.

\begin{figure*}[t]
\centering
\includegraphics[width=\textwidth]{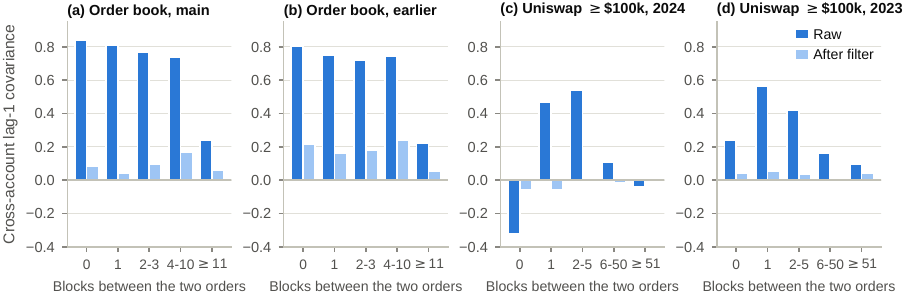}
\Description{Four bar charts of the cross-account lag-one covariance of order signs by block gap, raw and after filtering. In the two order-book panels the raw covariance is about 0.8 up to ten blocks apart and the filtered covariance is small. In the two Uniswap panels the same-block covariance is negative or weak, the raw covariance peaks one to five blocks apart, and filtering removes nearly all of it.}
\caption{Cross-account lag-one covariance of order signs by the number of blocks between the two orders, before (dark) and after (light) removing the prediction of the Binance-only filter. On the order book, the co-movement is strong within a block and up to ten blocks apart. Among large Uniswap swaps, it sits in the next few blocks, and within a block its sign is unstable.}
\label{fig:race}
\end{figure*}

\subsection{Where \texorpdfstring{$\kappa$}{kappa} is high, accounts race for the CEX price}
\paragraph{Timing.} Figure~\ref{fig:race}a,b shows the order book. The cross-account covariance of consecutive orders is 0.84 (main sample) and 0.80 (earlier sample) when both orders are in the same block, 0.81 and 0.75 in adjacent blocks, 0.72--0.77 two to ten blocks apart, and 0.24 and 0.22 beyond. Pairs at most ten blocks (a few seconds) apart contribute 90\% and 96\% of $C_{\mathrm{other}}(1)$, and same-block pairs alone contribute 17\% and 20\%. Orders in the same block were submitted before either could observe the other's fill, so their co-movement cannot be imitation of observed fills. Nor does the correlation rise once fills become observable.

Among large Uniswap swaps (Figure~\ref{fig:race}c,d), the co-movement sits one to five blocks apart (0.42--0.56), and these pairs contribute 94\% (2024) and 76\% (2023) of $C_{\mathrm{other}}(1)$. Within a block the covariance is negative in 2024 ($-0.32$) and weaker in 2023 (0.24). One interpretation is market design. In an automated market maker (AMM), every swap moves the pool price, so the first arbitrage swap of a block consumes much of the gap to the CEX price, and the race is decided block by block. On the order book, by contrast, several makers' quotes can be stale at the same time and several accounts take them within one block. There, the first and second same-direction orders in a block capture 2.11 and 2.01\bp{} relative to the CEX-implied fair value. Testing this interpretation directly is left for future work.

\begin{table}[t]
\caption{Learned common-signal filter. $R^2$: out-of-fold share of sign variance explained. Removed: share of the cross-account lag-one covariance removed by the filter, over all pairs, within the same block, and across adjacent blocks. Values above 100 mean the residual covariance is negative; ``--'' marks a negative raw covariance. Uniswap filters are fitted on swaps of at least \$10k.}
\label{tab:filter}
\small
\begin{tabular}{@{}llrrrr@{}}
\toprule
 & & & \multicolumn{3}{c}{Removed (\%)}\\
\cmidrule(l){4-6}
Sample & Features & $R^2$ (\%) & All & Same & Next\\
\midrule
Order book, main & Binance & 69 & 84 & 90 & 95 \\
\quad delayed 0.5 s & Binance & 68 & 83 & 84 & 94 \\
\quad with gap & Binance, gap & 73 & 87 & 92 & 96 \\
Order book, earlier & Binance & 54 & 73 & 73 & 78 \\
\midrule
Uniswap 2024, $\geq$ \$100k & Binance & 60 & 109 & -- & 112 \\
\quad with gap & Binance, gap & 66 & 89 & -- & 96 \\
Uniswap 2023, $\geq$ \$100k & Binance & 48 & 91 & 82 & 91 \\
\quad with gap & Binance, gap & 61 & 95 & 77 & 98 \\
\bottomrule
\end{tabular}
\end{table}

\paragraph{A learned filter removes most of the co-movement.} Table~\ref{tab:filter} reports the filter results. Using only Binance data, the filter explains 69\% of the variance of order-book signs in the main sample. It removes 84\% of the cross-account covariance: 90\% within blocks and 95\% across adjacent blocks. Delaying its inputs by 0.5\,s barely changes this (68\% and 83\%), and adding the venue--Binance gap raises the two figures to 73\% and 87\%. The earlier, less concentrated sample gives 54\% and 73\%. A residual cross-account correlation of 0.33 (main) and 0.40 (earlier) remains, against 0.63 and 0.69 before filtering, so other common signals, such as other leading venues or the venue's own state, may also matter.

For Uniswap swaps above \$100k, the Binance-only filter explains 48--60\% of signs and removes 91--109\% of the cross-wallet covariance; with the pool--Binance gap, the figures are 61--66\% and 89--95\%. The residual cross-wallet correlation lies between $-0.06$ and $0.08$. In these samples, then, most of the cross-account co-movement where $\kappa$ is high is a common response to the CEX price.

\begin{table}[t]
\caption{Adverse selection by flow type. Markouts at 60\,s (bp, dollar-weighted) of the liquidity side by dollar tercile of alignment $\alpha_n=\epsilon_n\hat s_n$ with the Binance-only filter (T1 least aligned), and for orders in a cross-account race (another account's same-direction order within three blocks on the order book, two on Uniswap) versus the rest. Order book: before trading fees and gas. Uniswap: swaps of at least \$10k; LP markouts include the 5\bp{} fee.}
\label{tab:econ}
\small
\setlength{\tabcolsep}{3.5pt}
\begin{tabular}{@{}lrrrrrr@{}}
\toprule
 & \multicolumn{3}{c}{Alignment tercile} & \multicolumn{3}{c}{Race}\\
\cmidrule(lr){2-4}\cmidrule(l){5-7}
Sample & T1 & T2 & T3 & \% of \$ & Race & Rest\\
\midrule
Book, main (makers) & $-$0.51 & $-$1.51 & $-$2.31 & 54 & $-$1.87 & $-$0.94 \\
Book, earlier (makers) & $+$0.08 & $-$1.73 & $-$2.48 & 74 & $-$1.70 & $-$0.46 \\
Uniswap 2024 (LPs) & $+$4.57 & $-$2.56 & $-$4.23 & 66 & $-$2.38 & $+$2.43 \\
Uniswap 2023 (LPs) & $+$5.29 & $-$2.33 & $-$5.08 & 78 & $-$2.44 & $+$5.57 \\
\bottomrule
\end{tabular}
\end{table}

\subsection{Race flow and adverse selection}
Table~\ref{tab:econ} reports the costs to liquidity providers. Sorting orders into dollar terciles of their alignment $\alpha_n=\epsilon_n\hat s_n$ with the filter's prediction, makers' markouts on the order book fall from $-0.51$ to $-2.31$\bp{} in the main sample and from $+0.08$ to $-2.48$\bp{} in the earlier one. Orders that belong to a cross-account race are 54\% and 74\% of volume. They cost makers 1.87 and 1.70\bp, and account for 70\% and 91\% of makers' losses; the remaining orders cost 0.94 and 0.46\bp.

On Uniswap, LP markouts after the 5\bp{} fee fall from $+4.6$ to $-4.2$\bp{} (2024) and from $+5.3$ to $-5.1$\bp{} (2023) across alignment terciles. Before the fee, LPs roughly break even on the least aligned third ($-0.4$ and $+0.3$\bp) and lose 9--10\bp{} on the most aligned third. Swaps in a cross-wallet race (66--78\% of volume) lose LPs about 2.4\bp{} even after the fee, while the rest earn 2.4--5.6\bp. The high-$\kappa$ race component of order flow is thus the adverse-selection component, the source of what the AMM literature calls loss-versus-rebalancing~\cite{milionis2022automated}.

\section{Discussion}
\paragraph{Reading DeFi stylized facts.} The departure that~\cite{dinosse2026deviations} report is real, but it describes the population of accounts as much as their behavior. Pair by pair, small swaps on Uniswap are dominated by splitting, as in equity markets~\cite{toth2015persistent,sato2023inferring}. What is new on-chain is a race among takers for the CEX price, in which orders from different accounts are nearly as correlated as orders from one account. Whenever venues or periods with different numbers of accounts are compared, we suggest reporting $\kappa$ (or $S$ relative to $\sum_iP_i^2$) next to the herding share.

\paragraph{Modeling.} Di Nosse et al.\ motivate order-level simulators of DEX dynamics~\cite{dinosse2026deviations}. Such simulators, and agent-based models of liquidity provision against latency-sensitive arbitrageurs~\cite{dinosse2026dynamicfees}, should combine traders who split their orders with arbitrageurs who respond to an exogenous CEX signal. They should also reproduce the venue-specific timing of the race.

\paragraph{Machine learning.} We use the learned filter as a measurement device. It isolates the predictable common component of order flow before the decomposition is applied, and cross-fitting keeps its own fit out of the residual statistics. Its output is also a pre-trade toxicity score whose terciles order liquidity providers' losses monotonically. The broader point concerns markets populated by algorithmic agents: agents that react to the same public signal generate herding statistics without imitating one another, so herding measures should be conditioned on the signals those agents observe.

\paragraph{Limitations.} We study one order-book venue and one contract over 188 hours, and one Uniswap pool over two weeks. An account is not a trader: one entity can spread its flow over several addresses, and routers or shared senders can merge many users. On the order book, merging accounts that share a sending address, which over-merges the users of shared gateways, raises $\kappa$ from 0.76 to 0.85; merging the two largest takers gives 0.73. Order-book markouts exclude trading fees and gas, and the delayed-input check covers the main sample only. The filter sees only Binance; other leading venues would likely explain more of the co-movement.

\section{Conclusion}
The herding share of order-flow persistence confounds the number of accounts with how strongly they co-move. Separating the two shows that small DEX flow is dominated by splitting pair by pair, and that much of the genuine cross-account co-movement in on-chain markets is a race for the centralized-exchange price. A learned filter removes most of that co-movement, and liquidity providers pay mostly for it.

\bibliographystyle{ACM-Reference-Format}
\bibliography{references}
\end{document}